\documentclass[runningheads]{llncs}
\usepackage[T1]{fontenc}
\usepackage{booktabs}
\usepackage{amsmath}

\usepackage{graphicx,verbatim}
\usepackage{amssymb}
\begin{document}
\title{impuTMAE: Multi-modal Transformer with Masked Pre-training for Missing Modalities Imputation in Cancer Survival Prediction}

\author{Maria Boyko\inst{1,2}  
\and Aleksandra Beliaeva \inst{1,2,3}
\and  Dmitriy Kornilov \inst{1,2}
\and Alexander Bernstein \inst{1}
\and Maxim Sharaev\inst{1,2}}

\institute{Center for Applied AI, Skolkovo Institute of Science and Technology, Moscow, Russian Federation \and
BIMAI-Lab, Biomedically Informed Artificial Intelligence Laboratory, University of Sharjah, Sharjah, United Arab Emirates
\and Ivannikov Institute for System Programming of the Russian Academy of Sciences, Moscow, Russian Federation}

\maketitle              
\begin{abstract}
The use of diverse modalities, such as omics, medical images, and clinical data can not only improve the performance of prognostic models but also deepen an understanding of disease mechanisms and facilitate the development of novel treatment approaches. However, medical data are complex, often incomplete, and contains missing modalities, making effective handling its crucial for training multimodal models. We introduce \textbf{impuTMAE}, a novel transformer-based end-to-end approach with an efficient multimodal pre-training strategy. It learns inter- and intra-modal interactions while simultaneously imputing missing modalities by reconstructing masked patches. Our model is pre-trained on heterogeneous, incomplete data and fine-tuned for glioma survival prediction using TCGA-GBM/LGG and BraTS datasets, integrating five modalities: genetic (DNAm, RNA-seq), imaging (MRI, WSI), and clinical data. By addressing missing data during pre-training and enabling efficient resource utilization, impuTMAE surpasses prior multimodal approaches, achieving state-of-the-art performance in glioma patient survival prediction. Our code is available at \url{https://github.com/maryjis/mtcp}


\end{abstract}
\section{Introduction}

Cancer is one of the most lethal and complex diseases \cite{who2024}, and requires a comprehensive approach to treatment and prognosis. Precision medicine aims to understand disease progression and identify patient-specific biomarkers for targeted therapy. Glioma, one of the most malignant and treatment-resistant brain tumors, is characterized by high heterogeneity and poor survival \cite{cancers3010621}. Accurate survival predictions help personalize treatments, enhance efficacy, and provide essential prognostic insights for patients and clinicians.

A major challenge in medical data is the limited availability of large datasets comparable to those in language models and image processing, where deep learning excels. Additionally, multimodal medical datasets are often incomplete, complicating their use in machine learning. For survival prognosis, MoME \cite{Xio_MoME_MICCAI2024} and PG-MLIF \cite{Pan_PGMLIF_MICCAI2024} fuse modalities but require complete data, support only four and two modalities, respectively, and lack essential pretraining. Pretrained methods \cite{zhou2023cross}, \cite{cheerla2019}, \cite{gomaa2024} improve cancer survival prediction but depend on contrastive learning, requiring all modalities during training, limiting usable samples. MultiSurv \cite{vale2020} avoids this but lacks pretraining. DRIM \cite{Rob_DRIM_MICCAI2024} pretrains encoders while reconstructing missing modalities but is computationally inefficient with six encoders.

Our method is based on the ViTMAE framework \cite{he2021}, which learns hidden representations by first masking a portion of the data and then attempting to reconstruct it using a decoder. Extensions of this method, such as MultiMAE \cite{bachmann2022multimae} and FLAVA \cite{hu2021}, have been proposed to enable its application to multiple modalities in the language-vision domain. In our study, we propose for the first time a multi-modal pretraining strategy for integrating five types of modalities related to different medical domains and training the decoder to predict the masked patches for all modalities simultaneously. This approach facilitates the learning of multimodal interactions during the pretraining stage. Unlike previous methods, which do not explicitly address missing modalities, our approach leverages available modality information to reconstruct missing components while learning a unified multimodal representation. More importantly, unlike contrastive learning, our method does not require all modalities to be present simultaneously. Instead, we can treat a missing modality as fully masked and train the decoder to reconstruct it. This allows for the effective utilization of all available medical data, which is crucial given the inherent scarcity of such datasets. This flexibility and computational efficiency make our method a powerful solution for multimodal medical data analysis, addressing the limitations of previous techniques. Thus, our contributions are:

\begin{enumerate}
\item We introduce an end-to-end unified multimodal approach that can be fine-tuned for downstream tasks and used to impute missing modalities.
\item  We apply a novel ViT-MAE-based pretraining strategy tailored for multimodal data with a high proportion of missing modalities.
\item Our proposed model exhibits greater universality compared to prior approaches and demonstrates scalability across an arbitrary number of modalities.
\item Our approach, being more universal in its integration of multiple modalities and robustness to missing data, achieving state-of-the-art performance on the glioma survival analysis task.
\end{enumerate}


\section{Methods}

ImpuTMAE is a multimodal end-to-end approach pretrained using a masked multimodal learning strategy, built upon a transformer architecture comprising a multimodal encoder and a multimodal decoder (Fig. 1). 

In the first stage, we pretrain the multimodal encoder-decoder to simultaneously reconstruct masked-out multimodal patches. Specifically, 50 \% of the patches from each of the four modalities (RNA, DNAm, WSI, MRI) were masked, and the model was trained to reconstruct them. Subsequently, in the second stage, we leveraged this pretrained model for a dual purpose: the decoder facilitates the reconstruction of missing modalities, while the encoder serves as a pretrained backbone for a downstream task—glioma survival analysis.

\begin{figure}
\includegraphics[width=\textwidth]{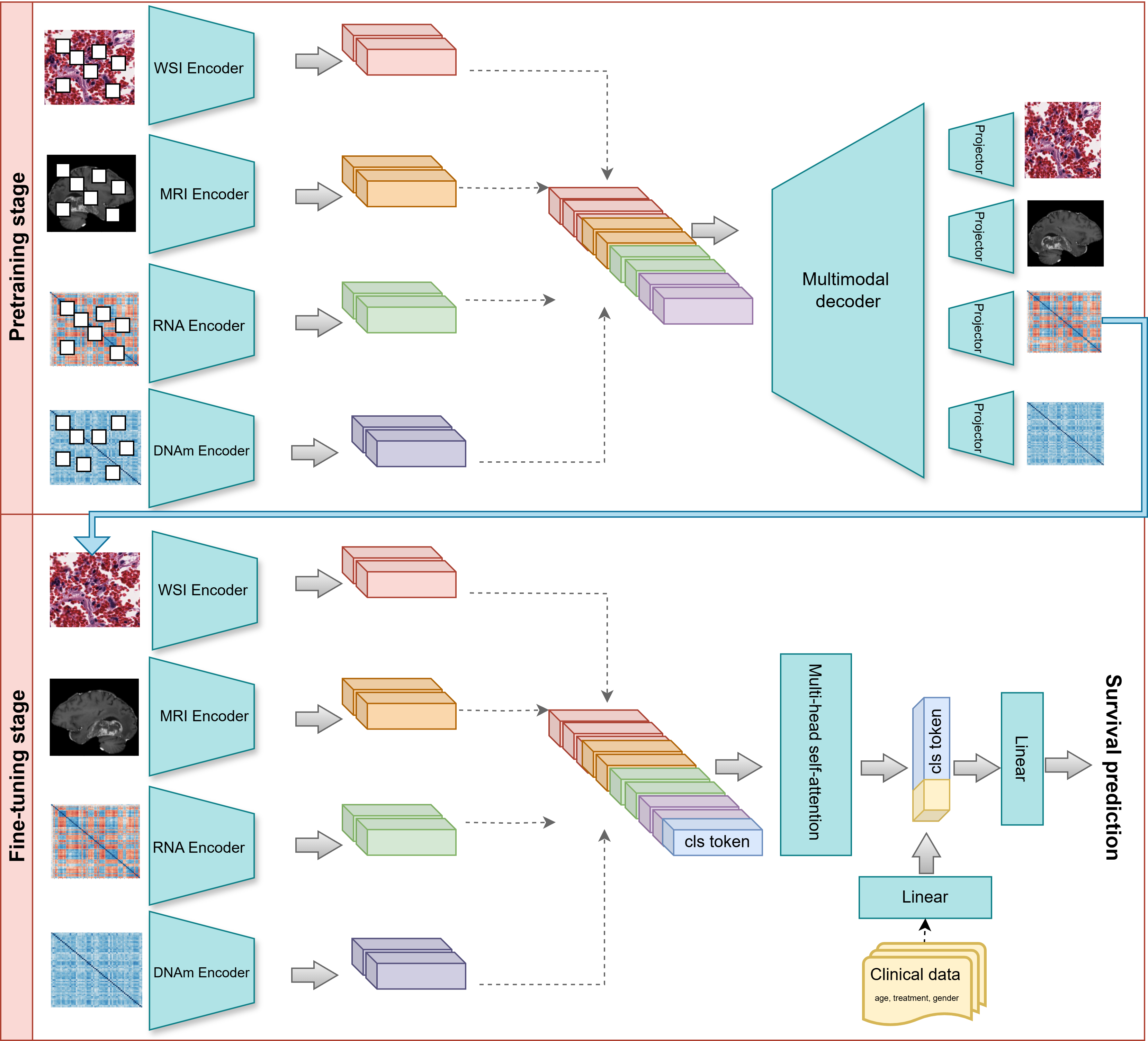}
\caption{An overview of ImputMAE architecture, which includes modality-specific encoders for DNAm, RNA, WSI, and MRI, as well as a multimodal decoder that processes the concatenated representation to simultaneously reconstruct masked-out multimodal patches. For glioma survival analysis, we integrate a fusion multi-head attention block into the pretrained encoder model and utilize the decoder for imputing missing modalities.} \label{fig1}
\end{figure}

\textbf{Multi-modal masked pre-training} 
While many previous approaches focus on reconstructing only a single modality or rely on separate encoders and decoders \cite{Rob_DRIM_MICCAI2024}, we propose a multimodal masked pre-training strategy, where a multimodal decoder jointly reconstructs patches from different modalities. Unlike methods that require using only the intersecting modalities for multimodal learning, we leverage all available multimodal data, including those with missing values, by treating missing modalities as entirely masked. The unmasked and non-missing modalities are processed through dedicated encoders for each modality, similar to ViT-MAE \cite{he2021}, while the masked patches and missing modalities are replaced with embeddings filled with \texttt{MASK\_TOKEN}.

We mask 50 percent of the patches from non-missing modalities and train the model to reconstruct them. Furthermore, the decoder reconstructs non-existent modalities, treating them as masked patches while omitting them from the loss computation. This pretraining strategy enables the model to learn multimodal representations that facilitate the imputation of missing modalities. As the reconstruction loss, we use the mean squared error (MSE) between the reconstructed and original data across different modalities. The loss is computed only on the masked patches of non-missing modalities. This strategy enables the model to learn robust representations that are resilient to missing data, effectively leveraging intermodal correlations for imputation.


 
\textbf{Modality-Specific Encoders}

Such medical data is inherently heterogeneous, with each modality (RNA-seq, DNA methylation, MRI, WSI) exhibiting a distinct structure. To effectively process this variability, each data modality is handled by a dedicated encoder network, which transforms the raw input into a latent feature embedding of fixed size (denoted as \(d\)). We design these encoders with architectures specifically tailored to their respective data types, incorporating modality-specific design choices such as the number of layers, layer types (e.g., fully connected or convolutional), output feature dimensions, and key hyperparameters (e.g., activation functions, normalization, and dropout strategies). This modular design enables each encoder to extract high-level features from its respective input modality before these representations are fused within a multimodal decoder

\textbf{RNA-Seq Encoder (Gene Expression)} The RNA-seq encoder segments the preprocessed RNA sequence into a series of non-overlapping patches of size 512 using a 1D convolution. Sinusoidal positional embeddings and a learnable CLS token are added to the patch embeddings. Subsequently, six transformer layers, similar to those in ViT-MAE, are applied to generate the hidden states  \(z_{\text{RNA}}\).

\textbf{DNA Methylation Encoder} follows a similar design to the RNA-Seq Encoder, consisting of a single convolutional layer that partitions the DNA methylation array into non-overlapping patches of size 1024. This is followed by six transformer layers, akin to those in ViT-MAE, to produce the hidden representations \(z_{\text{DNA}}\) .

\textbf{MRI Encoder (Imaging)} processes a volumetric MRI scan by dividing it into non-overlapping 3D patches of size 16x16x16. Each patch is embedded into a 256-dimensional vector via two 3D convolutions (kernel size and stride of 4). Four transformer layers, fewer than in RNA and DNA encoders due to quality saturation, are then applied to produce the hidden representations \(z_{\text{MRI}}\).

\textbf{WSI Encoder (Histopathology Image)} splits a 256$\times$256 patches into 16$\times$16 patches,  using a convolutional layer with a 16$\times$16 kernel and stride 16, resulting in the hidden representations \(z_{\text{WSI}}\). During pretraining, one random patch per sample is selected, while fine-tuning uses 10 patches per sample. The patch embeddings are then augmented with positional encodings and a CLS token before being processed by a transformer (4 layers, 4 attention heads).

 \textbf{Multimodal Decoder }
The outputs from the individual encoders are concatenated into fused latent representation [\(z_{\text{RNA}}\),\(z_{\text{DNA}}\), \(z_{\text{MRI}}\), \(z_{\text{WSI}}\) ] and fed to a multimodal decoder that reconstructs the original inputs for each modality. 


The multimodal decoder learns hidden representations through three transformer layers, similar to ViT-MAE, and then projects these representations back to the original patch size of each modality using dedicated projection heads. The projection head for WSI consists of a 2D transposed convolution layer, for MRI—a two 3D transposed convolution layers, and for RNA and DNA—linear layers. A reconstruction loss (mean squared error) is computed between the original data and the reconstructed data on the masked patches for each modality. The final multimodal MSE loss across all \(M\) modalities is computed as:
\begin{equation}
\text{MSE}_{\text{multimodal}} = \sum_{m=1}^{M} \text{MSE}^{(m)},
\end{equation}
\begin{equation}
\text{MSE}^{(m)} = \frac{1}{N_m} \sum_{i=1}^{N_m} \left(y_i^{(m)} - \hat{y}_i^{(m)}\right)^2, \quad m = 1, 2, \dots, M.
\end{equation}


\textbf{Fine-Tuning for Survival Analysis}

\textbf{Loading and Freezing Pretrained Modality-Specific Encoders:} We utilize pretrained encoders, which were jointly trained with the multimodal decoder, as feature extractors for fine-tuning. The MRI encoder and WSI encoder are fully frozen, while for both the DNAm and RNA encoders, five out of six transformer layers are frozen. 

\textbf{Handling Missing Modalities:} 
The pretrained multimodal decoder is employed to impute missing modalities, ensuring a consistent multimodal representation.

 \textbf{Fusion multi-head attention block:} The embeddings generated by each modality-specific encoder are concatenated into a unified multimodal representation. For WSI data, all patch embeddings are aggregated by averaging their CLS tokens before concatenation to construct the final patient-level representation. Finally, Fusion multi-head attention block is applied to effectively integrate heterogeneous features. This block consist of one self-attention block with 256 latent size.
 
 \textbf{Survival Risk Projection:}
The fused multimodal representation is processed through a final projection layer (a linear layer), which maps the multimodal embedding to survival risk scores across \(T\) time intervals. Following prior work \cite{Rob_DRIM_MICCAI2024}, we set \(T = 20\).

In survival analysis, we use a negative log-likelihood (NLL) loss based on hazard logits, defined as:

\begin{equation}
\mathcal{L} = - \frac{1}{n} \sum_{i=1}^{n} \sum_{t=1}^{\kappa(t_i)}
\left( y_{it} \log h(\tau_t | x_i) + (1 - y_{it}) \log(1 - h(\tau_t | x_i)) \right),
\end{equation}

where \( h(\tau_t | x_i) \) is the discrete-time hazard function, modeling the probability of an event at \( \tau_t \) given prior survival. The indicator \( y_{it} \) is 1 if the event occurs at \( t \), otherwise 0. The loss is averaged over \( n \) patients.

\section{Experiments}

We tested the proposed approach on multimodal glioma data from TCGA \cite{Chang2013}, focusing on the GBM and LGG cohorts, and incorporating five data modalities: DNA methylation (DNAm), RNA sequencing (RNA), histopathological whole-slide images (WSI), MRI scans—sourced from the BraTS dataset \cite{Baid2021} with selection limited to TCGA-listed patients. \textbf{RNA} data were filtered to retain protein-coding genes with variance >0.1, yielding 16,304 FPKM-UQ normalized genes, which were log-transformed. Gene expression was then normalized per subject, preserving inter-sample differences and reducing gene-specific noise more effectively than z-score normalization \cite{Bullard2010}. \textbf{DNA methylation (DNAm)} data (25,978 Beta values) were preprocessed following \cite{Rob_DRIM_MICCAI2024}. One artifact-free \textbf{WSI} slide per patient was manually selected. The full-resolution slide was subsampled into a thumbnail, and a tissue mask was extracted via the OTSU algorithm \cite{otsu1975}. From 1000 random 256×256-pixel patches, the 10 most informative—based on a custom HSV intensity score—were chosen for analysis.  Only T1-w \textbf{MRI} scans from BraTS \cite{Baid2021} were used. Tumor segmentation masks guided the extraction of a 64×64×64 voxel region, excluding healthy tissue. This region was resized, padded, and intensity-normalized for consistency in analysis. \textbf{Clinical} For clinical data, we selected three features: binary indicators for radiation treatment and pharmaceutical treatment, as well as patient age. We encoded the binary features with zeros/ones and applied z-score normalization to the continuous variable.


\textbf{Implementation Details} At the \textit{multimodal masked pretraining} stage, we trained ImputMAE on different modality subsets using the AdamW optimizer (\(\text{lr} = 1 \times 10^{-3}\), \(\text{wd} = 1 \times 10^{-2}\)) for 800 epochs with a cosine annealing scheduler and a batch size of 24. During the \textit{fine-tuning} stage for survival prognosis prediction, we trained the models for 20 epochs, applying a dropout rate of 0.1 for each modality subset while maintaining the same batch size, weight decay, optimizer, and scheduler as in the pretraining stage. The learning rate was set to \(1 \times 10^{-4}\) for RNA and DNA; \(1 \times 10^{-3}\) for RNA, DNA, and MRI; \(3 \times 10^{-4}\) for RNA, DNA, WSI and RNA, DNA, WSI, MRI.

\textbf{Evaluation} For evaluation, we split the dataset into 80\% training and 20\% testing sets. The best hyperparameters were selected via five-fold cross-validation, and models were assessed on the test set. We use the \textit{Concordance Index} (C-index) \cite{Antolini2005}, a widely used metric for ranking survival times, and the \textit{CS-score}, which provides a more objective assessment:
\begin{equation}
\text{CS-score} = \frac{\text{C-index} + (1 - \text{IBS})}{2}
\end{equation}
where IBS \cite{graf1999} is the Integrated Brier Score, reflecting calibration and discrimination. We evaluate multimodal models under two conditions:  
(1) \textbf{all modalities} available;  
(2) \textbf{at least RNA modality} present.

\section{Results}

\begin{table}[h]
    \centering
    \renewcommand{\arraystretch}{1.2}
    \setlength{\tabcolsep}{2pt}
    \begin{tabular}{l c c c c c c c c}
        \hline
        & \multicolumn{5}{c}{\textbf{Modality}} & \multicolumn{3}{c}{\textbf{Metrics}} \\
        \cline{2-6} \cline{7-9}
        & \textbf{rna} & \textbf{dna} & \textbf{MRI} & \textbf{WSI} & \textbf{cln} & \textbf{c-index} & \textbf{cs score} & \textbf{cs score*} \\
        \hline
        DRIM \cite{Rob_DRIM_MICCAI2024} & \textbf{\checkmark} &  \textbf{\checkmark}&  \textbf{\checkmark}& \textbf{\checkmark} &  & 0.825$\pm$0.012 & 0.856$\pm$0.009 & 0.863$\pm$0.049\\

        MultiSurv \cite{vale2020} & \textbf{\checkmark} & \textbf{\checkmark} &  & \textbf{\checkmark} & \textbf{\checkmark} & 0.790$\pm$0.027 & 0.711$\pm$0,021 & 0.880$\pm$0.051\\

        MoME \cite{Xio_MoME_MICCAI2024} & \textbf{\checkmark} & \textbf{\checkmark} &  & \textbf{\checkmark} &  & - & - & 0.696$\pm$0.028 \\
        \hline
        \textbf{imputMAE} & \textbf{\checkmark} &  &  &  &  & \textbf{0.828$\pm$0.009} & \textbf{0.864$\pm$0.004} & 0.864$\pm$0.004 \\
        \textbf{imputMAE} & \textbf{\checkmark} & \textbf{\checkmark} &  &  &  & \textbf{0.830$\pm$0.009} & \textbf{0.864$\pm$0.006} & 0.864$\pm$0.006 \\
        \textbf{imputMAE} & \textbf{\checkmark} & \textbf{\checkmark} & \textbf{\checkmark} &  &  & \textbf{0.834$\pm$0.008} & \textbf{0.865$\pm$0.002} &  \textbf{0.889$\pm$0.035}\\
        \textbf{imputMAE} & \textbf{\checkmark} & \textbf{\checkmark} &  & \textbf{\checkmark} &  & \textbf{0.834$\pm$0.006} & \textbf{0.867$\pm$0.003} & 
        0.858$\pm$0.006\\  
        \textbf{imputMAE} & \textbf{\checkmark} & \textbf{\checkmark} &  & \textbf{\checkmark} & \textbf{\checkmark} & \textbf{0.835$\pm$0.009}
        & \textbf{0.867$\pm$0.008} &
        0.857$\pm$0.011\\
        \textbf{imputMAE} & \textbf{\checkmark} & \textbf{\checkmark} & \textbf{\checkmark} & \textbf{\checkmark} &  & \textbf{0.827$\pm$0.006}
        & \textbf{0.869$\pm$0.002} &
        0.862$\pm$0.015\\
         \textbf{imputMAE} & \textbf{\checkmark} & \textbf{\checkmark} & \textbf{\checkmark} & & \textbf{\checkmark}  & \textbf{0.84$\pm$0.005}
        & \textbf{0.87$\pm$0.005} &
        \textbf{0.89$\pm$0.018}\\
        \textbf{imputMAE} & \textbf{\checkmark} & \textbf{\checkmark} & \textbf{\checkmark} & \textbf{\checkmark} & \textbf{\checkmark} & \textbf{0.831$\pm$0.004} & \textbf{0.864$\pm$0.004} & 0.863$\pm$0.008\\
        \hline
    \end{tabular}
    \caption{Performance comparison with previous approaches and across different modalities }
    \label{tab:results}
\end{table}

We compared our proposed method with previous state-of-the-art approaches, and the results are presented in Table 1. We evaluated model performance under two conditions: (1) all modalities available and (2) at least the RNA modality present. We selected these two evaluation settings because previous methods were primarily compared only in scenarios where all modalities were available (CS-score condition*). However, when evaluating methods on a large number of modalities (four or more), the number of samples containing all modalities decreases dramatically. To address this, in the second condition (C-index and CS-score), we evaluate our approach on the entire dataset where at least the RNA modality is available thus conduct more robust comparison. We impute missing modalities in prior approaches using currently proposed methods. We impute missing modalities with pretrained multimodal decoder. As we see in Table 1-2, our method consistently outperforms the vast majority of  prior works in both multimodal and unimodal settings across both conditions, demonstrating that a transformer-based approach with multimodal pretraining and missing modality imputation is one of the most effective solutions for survival prediction.

Additionally, our findings corroborate previous research \cite{vale2020},  \cite{Xio_MoME_MICCAI2024}, as shown in Table 2, confirming that RNA is the most critical modality for survival prediction. To further analyze the impact of individual components, Table 3 presents an ablation study evaluating the influence of the pretraining strategy on survival prediction performance. Thus, we demonstrate that the multimodal pretraining strategy and imputing missing modalities are useful steps for the downstream task. 

\begin{table}[t]
    \centering
    \renewcommand{\arraystretch}{1.2} 
    \setlength{\tabcolsep}{2pt} 
    \begin{tabular}{lcc}
        \toprule
        \textbf{} & \textbf{DRIM} & \textbf{imputMAE} \\
        \midrule
        RNA & 0.846$\pm$0.009 & \textbf{0.864$\pm$0.004} \\
        DNAm & 0.815$\pm$0.003 & \textbf{0.829$\pm$0.006}\\
        WSI & 0.776$\pm$0.007 &  0.708$\pm$0.024 \\
        MRI & 0.770$\pm$0.008 & \textbf{0.834$\pm$0.002} \\
        \bottomrule
    \end{tabular}
    \caption{Comparison of unimodal models performance using cs-score metric}
    \label{tab:example}
\end{table}

\begin{table}[t]
    \centering
    \renewcommand{\arraystretch}{1.2} 
    \setlength{\tabcolsep}{2pt} 
    \begin{tabular}{lcc}
        \toprule
        \textbf{} & \textbf{c-index} & \textbf{cs-score} \\
        \midrule
        imputMAE & 0.822$\pm$0.018 & 0.853$\pm$0.016 \\
        imputMAE +pretraining & \textbf{0.84$\pm$0.005} & \textbf{0.87$\pm$0.005}\\
        \bottomrule
    \end{tabular}
    \caption{Comparison contribution of multimodal pretraining strategy to survival analysis prognosis}
    \label{tab:example}
\end{table}

\section{Conclusion}
We propose an end-to-end unified multimodal framework with a multimodal pretraining strategy that can be fine-tuned for downstream tasks and leveraged to impute missing modalities. We demonstrate that this pretraining approach can effectively handle missing modalities and enable their imputation. These improvements allow our method to achieve state-of-the-art performance in the glioma patient survival prediction task. Moreover, the new multi-modal masked pre-training method ensures reliable model performance even with incomplete input data, which is especially important for clinical and biomedical applications where missing data is common
\clearpage
\bibliography{bib}
\bibliographystyle{splncs04}

\end{document}